\documentstyle[twocolumn,prl,eqsecnum,aps,epsf,12pt]{revtex}
\begin{document}
\tighten
\draft

\noindent

\title{\centerline{Properties of the Soliton-Lattice State}
       \centerline{in Double-Layer Quantum Hall Systems}}

\author{C. B. Hanna$^{(a)}$, A. H. MacDonald$^{(b)}$,
and S. M. Girvin$^{(b)}$}
\address{$^{(a)}$Department of Physics, Boise State University,
Boise, Idaho  83725\\
$^{(b)}$Department of Physics, Indiana University,
Bloomington, Indiana  47405\\}

\date{\today}

\maketitle

\begin{abstract}

\normalsize
\vspace{-0.3in}
Application of a sufficiently strong parallel magnetic field
$B_\parallel > B_{\rm c}$ produces a soliton-lattice (SL)
ground state in a double-layer quantum Hall system.
We calculate the ground-state properties of the SL state
as a function of $B_\parallel$ for total filling factor $\nu_{\rm T}=1$,
and obtain the total energy, anisotropic SL stiffness,
Kosterlitz-Thouless melting temperature, and SL magnetization.
The SL magnetization might be experimentally measurable,
and the magnetic susceptibility diverges as
$|B_\parallel - B_{\rm c}|^{-1}$.


\end{abstract}

\vspace{-0.1in}
\section{Introduction}
\label{sec:intro}
\vspace{-0.1in}

At sufficiently small layer separation, a double-layer quantum Hall
(2LQH) system\cite{dlexpt,dltheo} is an unusual quantum itinerant
ferromagnet.\cite{yang,moon}
The 2LQH system can be mapped to an equivalent spin-1/2
system by equating ``up'' (``down'') pseudospins with electrons
in the upper (lower) layer.\cite{yang}
For any finite layer separation $d$,
the itinerant ferromagnet has an $XY$ symmetry, so that
the orientation of a pseudospin at location ${\bf r}$ is
specified by its angle $\theta({\bf r})$ in the $xy$ plane.

Murphy {\it et al.} have investigated the effect of an in-plane
magnetic field $B_\parallel$ on 2LQH systems
and find evidence for a phase transition between two competing
QH ground states at a critical value
$B_\parallel = B_{\rm c}$.\cite{murphy}
These two ground states have been explained theoretically\cite{yang}
by showing that $B_\parallel$ produces a rotating Zeeman field
seen by the pseudospins.
This gives a pseudospin contribution to the ground-state energy
of the Pokrovsky-Talapov (PT) form,\cite{bak,dennijs}
\begin{equation}
E = \int d^2r \left [
\frac{\rho_{\rm s}}{2} |{\mathbf \nabla}\tilde{\theta} - {\bf Q}|^2 +
\frac{t}{2\pi\ell^2} (1 - \cos \tilde{\theta}) \right ] ,
\label{eq:tp2}
\end{equation}
where
${\bf Q}\equiv 2\pi{\bf \hat{z}\times B}_\parallel d/\phi_0$
defines the parallel magnetic-field wave vector,
$\tilde{\theta}({\bf r}) \equiv \theta({\bf r}) + {\bf Q\cdot r}$,
$t = t_0 e^{-Q^2\ell^2/4} \sqrt{4\nu_1\nu_2}$
is the tunneling energy ($t_0$ is the tunneling energy when
$Q=0$)\cite{hu} and
$\rho_{\rm s} = \rho_E (4\nu_1\nu_2)$ is the pseudospin stiffness
($\rho_E$ is the interlayer exchange stiffness when $\nu_1=\nu_2=1/2$)
in the Hartree-Fock Approximation (HFA),
$\nu_j$ is the filling factor of layer $j$,
and the energy is measured relative to
the ground-state energy for $B_\parallel = 0$.
Note that by adjusting the front and back gate voltages of the sample,
$\nu_1$ and $\nu_2$ may be varied (with $\nu_{\rm T}\equiv\nu_1+\nu_2=1$),
thereby allowing $t$ and $\rho_{\rm s}$ to be adjusted.

For small $B_\parallel$, Eq. (\ref{eq:tp2})
is minimized by $\tilde{\theta}({\bf r})=0$.
This is the commensurate (C) ground state.
For all finite $B_\parallel > B_{\rm c}$,
the pseudospin polarization has broken translational symmetry,
and a soliton-lattice (SL) state results.
For large $B_\parallel$
the pseudospins behave (almost) as if $t=0$.
This work focusses on calculating the ground-state properties of
the SL state, for all $B_\parallel > B_{\rm c}$.
Interestingly, it is not necessary to solve for the form of
$\tilde{\theta}({\bf r})$ in order to calculate the total energy of
the system.\cite{bak,dennijs,perring}

Minimizing $E$ with respect to
$\tilde{\theta}$ gives the 2D sine-Gordon equation (SGE),
$\xi^2 \nabla^2 \tilde{\theta} = \sin\tilde{\theta}$,
where $\xi/\ell = \sqrt{2\pi\rho_{\rm s}/t}$.
We shall give numerical values of our results
for a hypothetical GaAs 2LQH sample with
total density $1.1\times 10^{11} \rm{cm}^{-2}$,
layer separation $d=21.1$ nm,
and tunneling energy $t_0=0.1$ meV.
Such a sample would have $\ell\approx11.8$ nm, $d/\ell=1.8$, and
$\hbar\omega_{\rm c}\approx 8$ meV for $\nu_{\rm T}=1$,
and $\rho_{\rm E}\approx0.08$ meV
and $\xi\approx 26.5$ nm
in the HFA.

\vspace{-0.1in}
\section{Soliton-lattice state}
\label{sec:sls}
\vspace{-0.1in}

We investigate solutions of the SGE of the form
$\tilde{\theta}({\bf r}) =
\tilde{\theta}\left[{\bf\hat{e}}_1 \cdot ({\bf r}-{\bf r}_0)\right]$,
where ${\bf\hat{e}}_1$ is some unit vector in the $xy$ plane,
so that
$\xi^2\partial^2_1 \tilde{\theta} = \sin\tilde{\theta}$,
which is closely analogous to the equation of motion of
a pendulum.
The conserved quantity analogous to the total
energy of a pendulum is
$2 c^2 \equiv
(1/2) \xi^2 (\partial_1 \tilde{\theta})^2 -
(1 - \cos \tilde{\theta})$.
Defining $\beta=\tilde{\theta}/2$ leads to the equation
\begin{equation}
\xi\partial_1 \beta = \pm
\sqrt{c^2 +  \sin^2 \beta} .
\label{eq:beta}
\end{equation}
When $c=0$, Eq. (\ref{eq:beta}) gives
$\tilde{\theta}_{\rm ss}({\bf r}) = 4 \arctan \exp
[{\bf\hat{e}}_1\cdot ({\bf r} - {\bf r}_0)/\xi]$,
which represents a single soliton of width $\xi$,
corresponding to the motion of a pendulum that
completes a single revolution in infinite time.
The energy per unit length of a single soliton
relative to the C ground-state energy
may be computed from Eq. (\ref{eq:tp2}) as
$E_{ss}/L_2 = \rho_{\rm s} \left ( 8/\xi -
2\pi {\bf\hat{e}}_1 \cdot {\bf Q} \right )$,
where $L_2$ is the sample length perpendicular to ${\bf\hat{e}}_1$.
The lowest energy solitons have ${\bf\hat{e}}_1 = {\bf\hat{Q}}$
and their energy vanishes when $Q = Q_{\rm c} \equiv 4/(\pi\xi)$.
Thus for $Q>Q_{\rm c}$ it is energetically favorable to create solitons.
The solitons are weakly repulsive, and form a SL.
Analogous soliton effects also occur in long Josephson
junctions.\cite{stephen}

The SL spacing $L_{\rm s}$ may be determined by
noting that over one period of the SL,
$\beta$ changes by $\pi$.
Thus
\begin{equation}
L_{\rm s} =
\int_{-\pi/2}^{\pi/2} \frac{d\beta}{\partial_1\beta}
= 2\eta K(\eta) ,
\label{eq:ls}
\end{equation}
where we have used Eq. (\ref{eq:beta}), defined
$\eta \equiv 1/\sqrt{c^2 + 1}$, and where $K(\eta)$
is the complete elliptic integral of the first kind.\cite{gr}
We define the SL wave vector
${\bf Q}_{\rm s} \equiv (2\pi/L_{\rm s}){\bf\hat{e}}_1$,
so that Eq. (\ref{eq:ls}) becomes
\begin{equation}
Q_{\rm s}/Q_{\rm c} = (\pi/2)^2/[\eta K(\eta)] .
\label{eq:qsoqc}
\end{equation}
Note that $\eta \rightarrow 0$ corresponds to
$Q_{\rm s} \rightarrow Q \rightarrow \infty$,
whereas $\eta \rightarrow 1$ corresponds to the
C-SL transition, where $Q \rightarrow Q_{\rm c}$ and
$Q_{\rm s} \rightarrow 0$.

The energy per unit area is obtained by expressing Eq. (\ref{eq:tp2})
as an integral over $\beta$ [cf. Eq. (\ref{eq:ls})], which gives
\begin{eqnarray}
\frac{E}{L_1L_2} &=&
\frac{\rho_{\rm s}}{\xi^2} [
(\frac{Q^2}{2} - {\bf Q} \cdot {\bf Q}_{\rm s}) \xi^2
\\ \nonumber &-&
2(\frac{1}{\eta^2} - 1)
+ Q_{\rm c} Q_{\rm s} \xi^2 \frac{E(\eta)}{\eta} ] ,
\label{eq:elxly}
\end{eqnarray}
where $E(\eta)$ is the complete elliptic integral of the second
kind\cite{gr} and $L_1L_2$ is the sample area.
The value of ${\bf Q}_{\rm s}$ that minimizes the energy per unit area
is found by differentiating Eq. (2.4) with respect to
${\bf Q}_{\rm s}$, holding ${\bf Q}$ constant.
Using the identity\cite{gr}
$dE/d\eta = [E(\eta) - K(\eta)]/\eta$, one obtains
\begin{equation}
Q/Q_{\rm c} = E(\bar{\eta})/\bar{\eta} .
\label{eq:qoqc}
\end{equation}
Equations (\ref{eq:qsoqc}) and (\ref{eq:qoqc}) together
determine the equilibrium SL wave vector
$\bar{\bf Q}_{\rm s}({\bf Q})$ that minimizes the energy.
For $Q/Q_{\rm c}\rightarrow\infty$,
$\bar{Q}_{\rm s}/Q \approx
1 - (1/2)[(\pi/4) (Q_{\rm c}/Q)]^4$.
For $Q\rightarrow Q_{\rm c}$,
$\bar{Q}_{\rm s}/Q_{\rm c} \sim
(\pi^2/2)/\ln(1/\epsilon)$
asymptotically,\cite{bak,read}
where $\epsilon \equiv Q/Q_{\rm c}-1$.
If it were possible to achieve $\epsilon\sim 10^{-2}$
(e.g., by  gating the sample to tune $Q_{\rm c}$),
surface acoustic wave (SAW) techniques might detect the SL when
$L_{\rm s}$ matched the SAW wavelength.
The general SL ground-state solution for $\tilde{\theta}$
is given by
$\sin[(\tilde{\theta}_{\rm SL}-\pi)/2]
= {\rm sn}[{\bf\hat{e}}_1 \cdot ({\bf r}-{\bf r}_0)/\eta\xi,\eta]$,
where sn is the sine-amplitude Jacobian function.\cite{stephen}

\vspace{-0.1in}
\section{Stiffnesses}
\label{sec:stiffness}
\vspace{-0.1in}
 
The elastic constants of the soliton lattice are given by
the stiffness tensor
\begin{equation}
K_{ij} \equiv \lim_{{\bf Q}_{\rm s}\rightarrow\bar{\bf Q}_{\rm s}}
\left[ \frac{\partial^2 (E/L_1L_2)}
{\partial Q_{{\rm s}i} \partial Q_{{\rm s}j}}
\right]_{\bf Q} ,
\label{eq:kij} \end{equation}
which we calculate here for all $Q \ge Q_{\rm c}$
using Eqs. (2.4) and (\ref{eq:qoqc}).
Our results agree with those obtained in Ref. \onlinecite{read}
for $Q \rightarrow Q_{\rm c}$.

Using the results of Sec. \ref{sec:sls},
and the identity\cite{gr}
$dK/d\eta = E(\eta)/[\eta(1-\eta^2)] - K(\eta)/\eta$,
the compressional elastic constant $K_{11}$ is found to be
\begin{equation}
\frac{K_{11}}{\rho_{\rm s}} = 
\frac{\partial Q}{\partial \bar{Q}_{\rm s}}
\rightarrow \left \lbrace
\begin{array}{cc}
1 - 
\frac{3}{2} \left ( \frac{\pi}{4} \frac{Q_{\rm c}}{Q} \right )^4 ,
& Q/Q_{\rm c} \rightarrow \infty \\
(2/\pi^2) \epsilon \ln^2 (1/\epsilon) ,
& Q/Q_{\rm c} \rightarrow 1
\end{array} \right .
\label{eq:k11}
\end{equation}
As expected, $K_{11}\rightarrow\rho_{\rm s}$
in the limit $Q/Q_{\rm c} \rightarrow \infty$;
$K_{11}\rightarrow 0$ for $Q/Q_{\rm c} \rightarrow 1$
because of the short-range repulsions between the solitons.
The shear elastic constant $K_{22}$ is
\begin{equation}
\frac{K_{22}}{\rho_{\rm s}} = \frac{Q}{\bar{Q}_{\rm s}}
\rightarrow \left \lbrace
\begin{array}{cc}
1 + 
\frac{1}{2} \left ( \frac{\pi}{4} \frac{Q_{\rm c}}{Q} \right )^4 ,
& Q/Q_{\rm c} \rightarrow \infty \\
(2/\pi^2) \ln (1/\epsilon) ,
& Q/Q_{\rm c} \rightarrow 1
\end{array} \right .
\label{eq:k22}
\end{equation}
Like $K_{11}$, approaches $K_{22}\rightarrow \rho_{\rm s}$
for $Q/Q_{\rm c} \rightarrow \infty$.
However, $K_{22}$ diverges for $Q\rightarrow Q_{\rm c}$.

The SL phase of the PT model can undergo a KT
transition,\cite{dennijs}
with a transition temperature
$k_B T_{\rm KT}\sim(\pi/2)\sqrt{K_{11}K_{22}}$.
The KT transition is probably most easily measured in devices
with oppositely directed currents in each layer.\cite{moon}
However, because the SL dislocations are electrically charged,
the KT transition might increase the longitudinal resistivity
even in devices without separately contacted layers,
due to the proliferation of unbound charged dislocations above $T_{\rm KT}$.

We now calculate the stiffness tensor $\tilde{K}_{ij}$
relevant to the sound velocities of the SL by writing
$\tilde{\theta}({\bf r})=
\tilde{\theta}_0({\bf r})+\delta\tilde{\theta}({\bf r})$,
where $\tilde{\theta}_0$ is the ground-state solution
of the SGE which minimizes the PT energy (\ref{eq:tp2})
and $\delta\tilde{\theta}$ is a small deviation of $\tilde{\theta}$
from $\tilde{\theta}_0$.
To quadratic order in $\tilde{\theta}$, the change in the PT energy
(\ref{eq:tp2}) of the 2LQH system is
\begin{equation}
\delta E =
\frac{1}{2} \int \frac{d^2r}{2\pi\ell^2}
\delta\tilde{\theta}
\left[ t\cos\tilde{\theta}_0 -
2\pi\ell^2 \rho_{\rm s} \nabla^2 \right]
\delta\tilde{\theta}
\label{eq:delh}
\end{equation}
where we have integrated by parts and made use of the SGE.
The energy $\delta E$ is minimized by choosing $\delta\tilde{\theta}$
from among the eigenvalues of bracketed Schr\"odinger-like
operator in Eq. (\ref{eq:delh}).
For ${\bf Q}={\bf \hat{x}}$,
$\tilde{\theta}_0=\tilde{\theta}_0(x)$
and we may write
$\delta\tilde{\theta}({\bf r})
\propto \exp(iq_yy) \delta\tilde{\theta}(x)$.
When $\tilde{\theta}_0({\bf r})=\tilde{\theta}_{SL}({\bf r})$,
the eigenvalue equation for bracketed equation in
Eq. (\ref{eq:delh}) becomes Lam\'e's equation after a simple
rescaling of $x$.\cite{stephen}
The relevant ``vortex oscillation'' solutions to Lam\'e's equation
follow from Ref. \onlinecite{stephen} and Eqs. (\ref{eq:qsoqc}) and
(\ref{eq:qoqc}) in the long-wavelength limit:
$\delta E/L_1L_2 = (1/2) (K_{11} q_x^2 + \rho_{\rm s} q_y^2 )$,
where $K_{11}$ is equal to the compressional stiffness
(\ref{eq:k11}).
Thus $\tilde{K}_{11}=K_{11}$, while
$\tilde{K}_{22}=\rho_{\rm s}$ is independent of $Q$.

\vspace{-0.1in}
\section{Magnetization and Magnetic Susceptibility}
\vspace{-0.1in}

The SL makes a small but possibly measurable contribution to
the magnetization of the 2LQH system.
The parallel-field magnetization per unit area is
\begin{equation}
{\bf M}_\parallel \equiv
-\frac{\partial}{\partial {\bf B}_\parallel}
\left( \frac{\bar{E}}{L_1L_2} \right) =
-M_0 ({\bf Q} - \bar{{\bf Q}}_{\rm s})/Q_{\rm c} ,
\label{eq:mag}
\end{equation}
where the equilibrium  value of $\bar{E}/L_1L_2$,
found by using Eqs. (\ref{eq:qsoqc}) and (\ref{eq:qoqc}) in Eq. (2.4),
and we have used the results of Sec. \ref{sec:sls}.
$M_0\equiv 2\pi\rho_{\rm s} Q_{\rm c}d/\phi_0$ sets the scale of the
magnetization density, which we plot in Fig. 1.
It is useful to compare the total SL magnetization
$M_0L_1L_2$ to the scale of the Landau diamagnetism in a $\nu=1$
QH system:
\begin{equation}
\frac{M_0L_1L_2}{N\mu_{\rm B}^*} =
\frac{M_0\phi_0}{\mu_{\rm B}^* B_\perp} =
16 \frac{d}{\xi} \frac{\rho_{\rm s}}{\hbar\omega_{\rm c}}
\sim 0.1 ,
\label{eq:m0}
\end{equation}
where $\mu_{\rm B}^* = e\hbar/2m^*c$ is the effective Bohr magneton in GaAs,
and $m^*\approx 0.067m_{\rm e}$ is the effective mass.
This shows that in the HFA, the SL magnetization is roughly
an order of magnitude smaller than the Landau diamagnetism,
which has been measured in GaAs single-layer
heterostructures.\cite{eisentorque}
The torque on a tilted 2LQH sample has both a Landau-diamagnetic component
$\tau_\perp = M_\perp B_\parallel$ and a SL component
$\tau_\parallel = M_\parallel B_\perp$.
The smallness $M_\parallel$ is somewhat compensated
by the size of $B_\perp$ in the expression for $\tau_\parallel$.

The parallel-field magnetic susceptibility
$\chi_\parallel \equiv \partial M/\partial B_\parallel$
is given by
\begin{eqnarray}
\chi_\parallel & = &
\chi_0 (\partial \bar{Q}_{\rm s}/\partial Q - 1) =
\chi_0 (\rho_{\rm s}/K_{11} - 1)
\\ \nonumber & \rightarrow &
\chi_0 \left \lbrace
\begin{array}{cc}
(3/2) [(\pi/4) (Q_{\rm c}/Q)]^4 ,
& Q/Q_{\rm c} \rightarrow \infty \\
(\pi^2/2) / [\epsilon \ln (1/\epsilon)] ,
& Q/Q_{\rm c} \rightarrow 1 \\
-1 ,
& Q < Q_{\rm c}
\end{array} \right .
\label{eq:chi}
\end{eqnarray}
where
$\chi_0 \equiv (2\pi d/\phi_0)^2 \rho_{\rm s}
\sim 5\times10^{-14}$m.
Near the C-SL transition, $\chi_\parallel \sim 1/\epsilon$,
 with logarithmic corrections.
It might not prove practical to make AC measurements of $\chi_\parallel$
because of substantial eddy-currents in the QH regime.

\vspace{-0.1in}
\section{Acknowledgments}
\vspace{-0.1in}

This work was supported by an award from Research Corporation,
and by the National Science Foundation under grant DMR94-16906.

\addcontentsline{toc}{part}{Figure Captions}

\vspace{-0.04in}
\noindent
FIG. 1.
SL contribution to the magnetization density (dots),
which drops precipitously for $Q>Q_{\rm c}$.
The C-phase result (solid line) holds for $Q<Q_{\rm c}$.


\begin{figure}[b]
\epsfxsize3.2in
\centerline{\hspace{0.3in}\epsffile{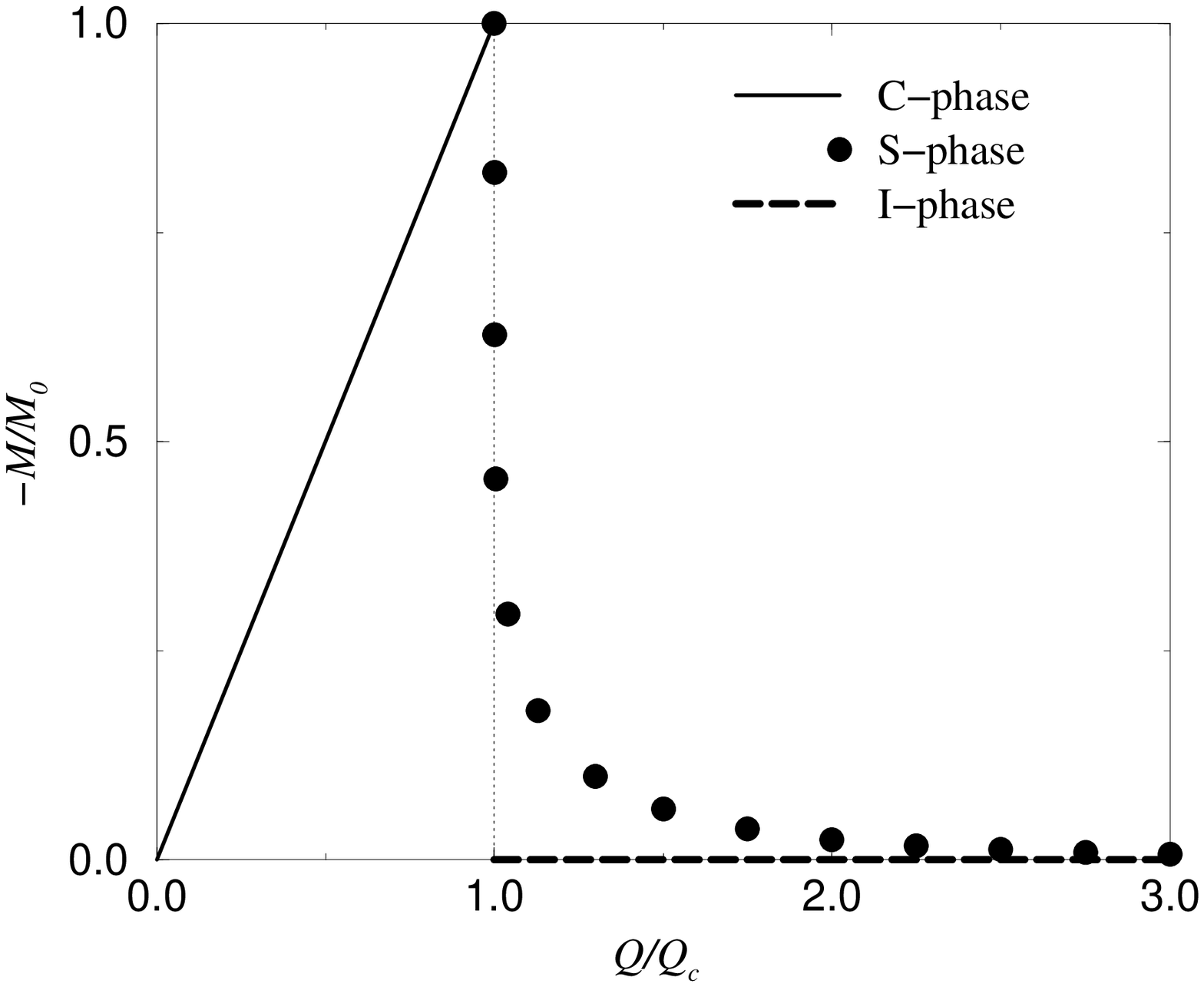}}
\end{figure}

\end{document}